\documentclass[aps,prl,twocolumn,groupedaddress,showpacs,floatfix]{revtex4}
\usepackage{graphicx}
\begin{document}
\title{Stress relief as the driving force for self-assembled Bi nanolines}
\author{J.H.G.Owen}
\email[]{james.owen@aist.go.jp} 
\author{K.Miki}
\email[]{miki.kazushi@aist.go.jp} 
\affiliation{Nanotechnology Research Institute, National Institute of
Advanced Industrial Science and Technology (AIST), AIST Tsukuba
Central 2, 1-1-1 Umezono, Tsukuba. Ibaraki 305-8568, Japan}
\author{H.Koh} 
\author{H.W. Yeom} 
\email[]{yeom@phya.yonsei.ac.kr}
\affiliation{Atomic-scale Surface Science Research Center and
Institute of Physics and Applied Physics, Yonsei University, Seoul
120-749, Korea}
\author{D.R.Bowler} 
\email[]{david.bowler@ucl.ac.uk}
\affiliation{Department of Physics and Astronomy, University College
London, Gower Street, London WC1E 6BT, UK}

\date{\today}

\begin{abstract}
Stress resulting from mismatch between a substrate and an adsorbed
material has often been thought to be the driving force for the
self-assembly of nanoscale structures. Bi nanolines self-assemble on
Si(001), and are remarkable for their straightness and length -- they
are often more than 400 nm long, and a kink in a nanoline has never
been observed. Through electronic structure calculations, we have
found an energetically favourable structure for these nanolines that
agrees with our scanning tunneling microscopy and photoemission
experiments; the structure has an extremely unusual subsurface
structure, comprising a double core of 7-membered rings of
silicon. Our proposed structure explains all the observed features of
the nanolines, and shows that surface stress resulting from the
mismatch between the Bi and the Si substrate are responsible for their
self-assembly.  This has wider implications for the controlled growth
of nanostructures on semiconductor surfaces.
\end{abstract}

\maketitle

Nanowires are of enormous importance for nanoelectronics: recently,
various devices have been constructed from semiconductor
nanowires\cite{Huang2001} and carbon nanotubes\cite{Bachtold2001}, to
name but two.  These approaches require assembly \textit{on} the
surface, whether by fluidics and patterning or use of scanning probes;
self-assembled nanowires would be a compelling alternative for
fabricating a large number of devices.  However, the microscopic
understanding of the underlying physical and chemical mechanisms for
self-assembly of nanoscale features has been limited. Surface stress
resulting from lattice mismatch in heteroepitaxial growth has often
been thought to be responsible for self-assembly of nanoscale features
(such as the growth of Ge "hut'' clusters\cite{Eaglesham1990}).  For
the case of self-assembled nanowires on semiconductor surfaces, there
has been much recent work on rare-earth silicides (e.g. ErSi$_2$),
where there is a large strain along one axis ($\sim7\%$) and almost
none along another, leading to the formation of extended
nanowires\cite{Chen2000}. These wires, however, are far from being
uniform or perfect at the atomic scale.  In contrast, Bi nanolines,
formed when a Bi-covered Si(001) surface is annealed at around
$570-600^\circ$C\cite{Miki1999a,Miki1999b}, are quite striking in
their uniformity. These nanolines are always 1.5 nm wide, and extend
for hundreds of nanometres without a kink or a defect.  As well as
this, they repel defects and down B-type step edges to a distance of
3-4nm.  Further, they are resistant to attack by radical hydrogen or
oxygen (hence hydrogen can be used as a mask and oxygen used to
isolate them electrically from the substrate)\cite{Owen2002}, making
them promising as templates for nanowires of other materials. However,
the structure of these Bi nanolines, a prerequisite to a microscopic
understanding of their unique properties and hence control of their
nucleation and growth, remains unknown.

Earlier, we proposed a structure based upon a 3-dimer wide
model\cite{Miki1999a,Bowler2000}. However, recent atomic resolution
scanning tunneling microscopy (STM) images of the Bi
nanoline\cite{Owen2002,Naitoh2000}, have revealed that the structure
in fact occupies the space of 4 dimers in the Si(001) surface. A
proposed four dimer model\cite{Naitoh2000} has the wrong spacing of
features in the nanoline (ca. 5.4 \AA), and is energetically very poor
(more than 0.6 eV/Bi dimer worse than the 3-dimer
model\cite{Bowler2000}).  Moreover, neither model has a large kinking
energy.  Accordingly, we have conducted an exhaustive search for a new
structure and tested many candidates against experimental criteria,
which we detail below.  For this purpose, semi-empirical tight binding
(tb) calculations are invaluable, as they allow us to run large
simulations of hundreds of atoms quickly on modest hardware(a desktop
computer), while achieving relaxed energies very close to those
obtained from \textit{ab initio} density functional theory (DFT)
calculations\cite{Bowler2000,Bowler2002}.  In this paper, we propose a
new structure for these nanolines that is energetically favourable and
agrees with all experimental observations (size, stability, registry
with Si dimers, straightness, repulsion of defects).  We note that it
is related to a structure recently proposed for B-type steps on
As-terminated Ge(001) surfaces, and that our findings may well have
implications for group V elements on group IV surfaces in general.

\begin{figure}
\includegraphics[width=\columnwidth]{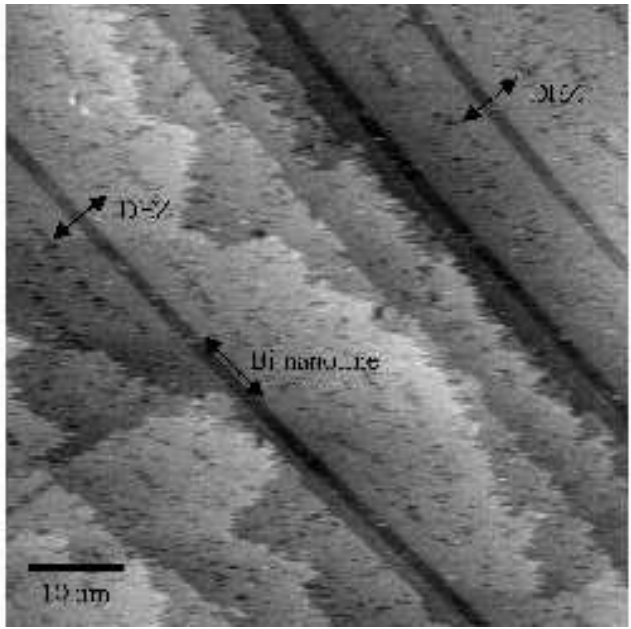}
\caption{A 65nm $\times$ 65nm STM image of the Si(001) surface taken
at 590$^\circ$C, showing 3 Bi nanolines. The black streaks on the
surface are rapidly moving defects.  Around each nanoline, there is an
area free of defects marked by the arrows as "DEZ" (Defect Exclusion
Zone). At this sample bias, the nanolines appear dark, at higher
biases, the nanoline is bright relative to the silicon.
\label{fig:dez}}
\end{figure}

The Si(001) substrate was cleaned using a standard
process\cite{Miki1998} before being transferred into vacuum. The Si
surface was prepared by flashing repeatedly to 1100$^\circ$C for a few
seconds, until there was only a small pressure rise. The clean surface
was checked with STM before Bi deposition began. Bi was evaporated
from an effusion cell, a typical dose being Bi at 470$^\circ$C for 10
mins. STM images were taken at the deposition temperature between
570-600$^\circ$C, and at room temperature, using a JEOL 4500 XT UHV
STM.  The high-resolution Bi 5d core-level photoemission spectra,
taken at 65 eV photon energy, were measured on the new high-resolution
vacuum ultraviolet beam line BL-1C at Photon Factory, KEK, Japan. The
overall energy resolution was better than 100 meV.  The relative
stabilities of the proposed structures were calculated using two
electronic structure techniques: for a swift search of possible
motifs, tight binding; for accurate energies and structures, density
functional theory(DFT). The tight binding calculations were performed
with a linear scaling code (an implementation of the Density Matrix
Method\cite{Li1993}) using a parameterisation which has been
previously validated for Bi-Si interactions\cite{Bowler2000}; this
allowed the large unit cells necessary to model the long range strain
effects seen in our STM data.  The DFT calculations were performed
using the VASP code\cite{Kresse1996}, using ultrasoft
pseudopotentials, a plane wave cutoff of 150 eV (sufficient for energy
difference convergence) and a Monkhurst-Pack \textbf{k}-point mesh
with 4$\times$4$\times$1 points.  The unit cell used had ten layers of
Si, with sixteen atoms in each layer (forming a single dimer row with
the p(2$\times$2) reconstruction) with the bottom two layers
constrained to remain fixed and dangling bonds terminated in hydrogen.
When comparing energies with different amounts of Bi, we use unit
cells of the same surface area, and compare the excess surface energy
plus bismuth adsorption energy per Bi dimer\cite{Bowler2000}.

\begin{figure}
\includegraphics[width=\columnwidth]{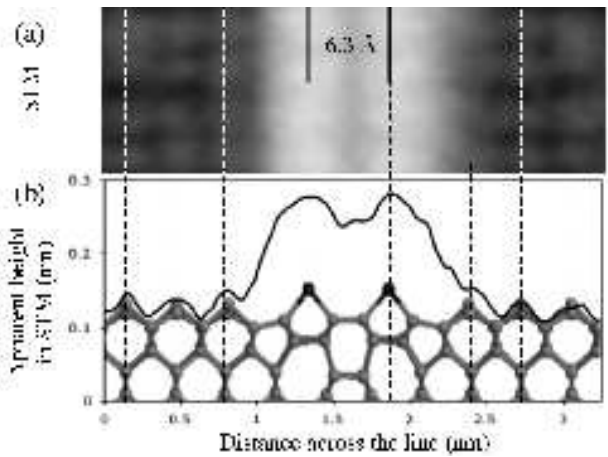}
\caption{(a): A Bi nanoline (on a H-terminated surface). The feature
spacing within the nanoline is 6.3\AA.  (H-termination of the
background Si dimers makes them easier to resolve, so that the
registry of the line relative to the surface can be confirmed.)  (b):
The side view of our proposed structure has been matched up to the STM
cross-section. Dotted lines mark the peak positions, showing that the
model and the STM match extremely well.
\label{fig:STM}}
\end{figure}

\begin{figure}
\includegraphics[width=\columnwidth]{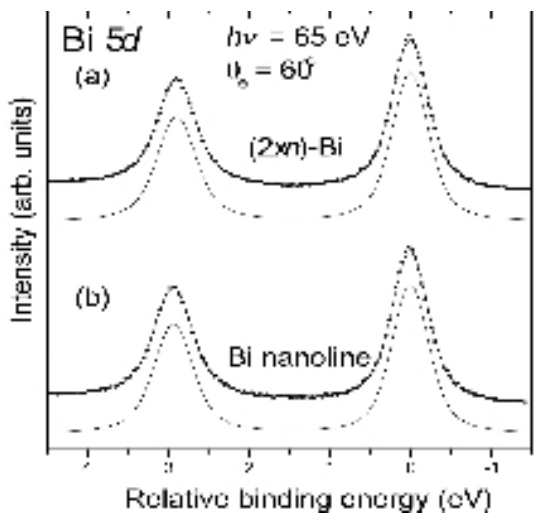}
\caption{The Bi 5d core-level photoemission spectra of the (a)
(2$\times$n)-Bi phase, (b) Bi nanoline phase. The raw data are dots
and the fitting curves are solid (dotted) lines.  The two curves are
essentially identical, suggesting that the local bonding of the Bi in
the nanoline and in the (22$\times$nn) phase are the same.
\label{fig:pes}}
\end{figure}

The Bi nanolines have several notable and unusual features. First,
their straightness and perfection.  Hundreds of lines have been
observed, many over 400 nm long, without a kink being seen, and
defects are extremely rare (their straightness can be seen in
Fig.~\ref{fig:dez} and also in previous
work\cite{Miki1999a,Miki1999b}). This would suggest that the nanoline
has a large kinking energy.  Second, the ``defect exclusion zone''
(DEZ).  Low concentrations of Bi embedded in the top layer of Si(001)
cause compressive surface stress, and ordered trenches of missing
dimer defects form every 8-12 dimers to relieve the stress.  Despite
being highly mobile at high temperatures, these defects do not come
within 3-4 nm of the nanolines -- the DEZ.  Since the strain field of
a missing dimer defect is tensile, the repulsive interaction between
the defect and the nanolines suggests that the nanoline strain field
should also be tensile, and hence is also a source of stress relief
for the embedded Bi. It is likely therefore that the stress induced by
the Bi in the top layer of the Si is the driving force for the
formation of this unusual structure.  Having formed, the nanolines
remain after epitaxial islands of Bi have evaporated, indicating
increased stability (RHEED experiments found the difference in
desorption barrier to be 0.25eV\cite{Miki1999a}). However, the local
structure of the Bi in the nanoline appears to be in a simple
dimerized form, in the top surface layer.  Recent high resolution STM
images, such as in Fig.\ref{fig:STM} (a), and previous
work\cite{Naitoh2000,Owen2002} show that the nanolines are 4 dimers or
15.4\AA wide. The bright dimer-like features making up the nanoline
lie between the underlying Si dimers, in line with the Si dimer rows.
As marked, the spacing between the nanoline features is 6.3 \AA.
Photoemission spectroscopy experiments, shown in Fig.\ref{fig:pes},
find that the Bi 5d core-level spectra of the Bi nanowire is
essentially identical to the spectra of the (2$\times$n) phase
composed of Bi ad-dimers, with a single well defined spin-orbit
doublet. This suggests that the local chemical state and registry of
Bi adsorbates for both phases are the same, i.e. that the Bi is in the
form of dimers in the top layer of the structure. X-ray photo-electron
diffraction (XPD) experiments\cite{Shimomura2000} confirm the presence
of Bi dimers parallel to the Si dimers and find the spacing between
them to be $\sim$6.3\AA. Hence, the observed properties of the Bi
nanoline must result not from a novel Bi structure, such as a square
of Bi atoms, but from an unusual Si substructure, stabilised by the
presence of Bi.


\begin{figure}
\includegraphics[width=\columnwidth]{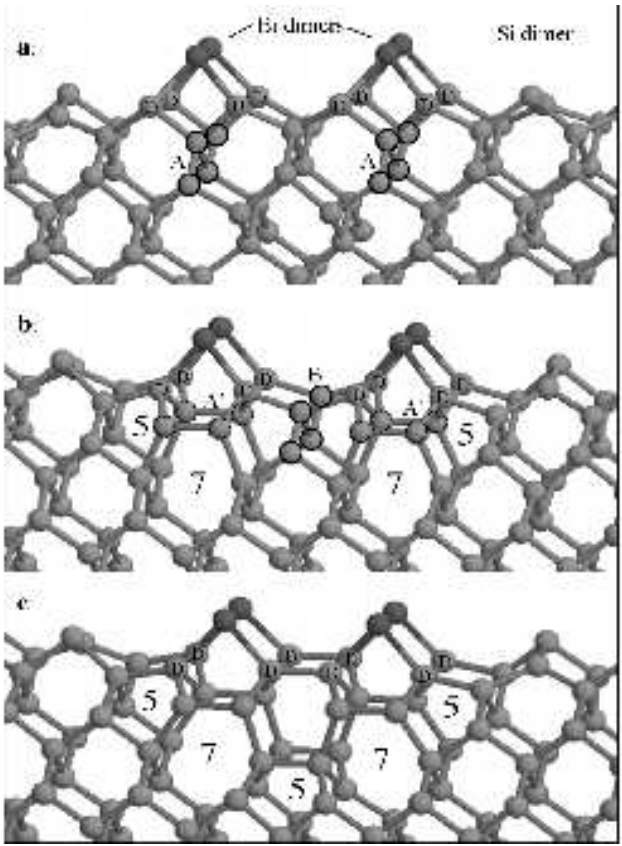}
\caption{Candidate Bi nanoline structures. In (a), a simple ad-dimer
structure is shown. The atoms marked 'D' are the original Si dimer
atoms. Rotation of the 2nd/3rd layer atoms (marked A) beneath the Bi
dimers, produces the second structure (b), with two cores of 5- and
7-membered rings of Si. This structure is is under compressive stress,
which may be relieved by the removal of the central four 2nd/3rd layer
atoms (marked B) and rebonding, resulting in our proposed structure
(c).  Our proposed structure has an energy 0.25 eV/dimer better than
the high coverage Bi-(2$\times$n) surface.
\label{fig:struc}}
\end{figure}

A simple model for the line, with two Bi ad-dimers sitting on top of
the dimer row, but between two Si dimers, as shown in
Fig.~\ref{fig:struc}(a), reproduces most of the aspects of the
detailed STM data, and is consistent with the PES and XPD data, but is
0.53 eV/Bi dimer less stable than the (2$\times$n) phase, and has no
energetic reason to grow long and straight; nor, indeed, is there any
reason for the two ad-dimers to remain adjacent. However, our proposed
structure may be reached from this simple structure by rearrangement
of only a few atoms, illustrated in Fig.~\ref{fig:struc}.  A simple
2-dimer wide core (equivalent to half of our proposed structure) can
be formed by rotation of the 2nd/3rd layer atoms, so that they lie on
the same level. The resulting 2-dimer structure is quite similar to
that proposed for the structure of As-Ge double-height B-type
steps\cite{Zhang2001}.  The energy of this structure is lower than the
simple ad-dimer model, but the strain field of the core is
compressive, the wrong sign for stress relief, and two such cores
close together, necessary to form a 4-dimer unit, as shown in
Fig.~\ref{fig:struc}(b), repel each other.  Removal of four central
atoms and rebonding of 1st and 4th-layer Si atoms makes the overall
strain field tensile, while keeping all bonds saturated (The
terminating species is important; replacement of the two Bi ad-dimers
by Si ad-dimers raises the total energy by $\sim$2eV/dimer).  This
gives our new proposed structure, shown in Fig.~\ref{fig:struc}(c).

Our proposed structure is energetically very favourable and gives good
agreement with all aspects of our experimental findings. In DFT
calculations, the energy/Bi dimer is -10.9 eV/Bi dimer, 0.25 eV/Bi
dimer lower than the high coverage Bi-($2\times$n) phase. This energy
difference agrees very well with the difference in stability as
measured by RHEED. Comparison of the structure to an STM profile
yields extremely good agreement. A cross-section of the nanoline
matched up with the ball-and-stick model is shown in
Fig.~\ref{fig:STM}(b).  The position of the peaks of the Bi dimers
line up perfectly with the STM profile. The spacing found from our
calculations (6.25\AA) and the direction of the Bi dimers (parallel to
Si dimers) agree extremely well with photoemission spectroscopy and
with XPD results\cite{Shimomura2000}.  Moreover, our proposed
structure stands out from others that we have tested, in that it
explains the other observed features of the Bi nanolines very
naturally. The importance of surface strain relief in the formation of
this structure is underlined by its increased stability in a surface
terminated by Bi dimers. In this case, the total energy of a long
tightbinding unit cell was further lowered by $\sim$0.1 eV/dimer, due
to relaxation of the compressive stress in the surface Bi dimers.  As
expected, the tensile strain field of the nanoline also provides a
driving force for the DEZ. Tightbinding calculations on a 32-dimer
cell found that single missing dimer defects and step edges interact
repulsively with the nanoline out to a range of $\sim$3 nm, in
agreement with the observed DEZ width of 3-4 nm.  The extensive
distortion down to the fifth layer leads to a large kinking energy
(3.75 eV/kink (tb)) and destabilises the dimers adjacent to the end of
the line. The excess energy is 2.6 eV/line end in tightbinding
calculations for an isolated line segment. We plan to report detailed
calculations on all these features of the nanolines in future work.

In conclusion, we have found an extraordinary structure for the Bi
nanoline, which involves extensive reconstruction down to the fifth
layer below the surface.  Not only is this structure more stable than
any other we have calculated, and matches extremely well with criteria
drawn from a variety of experimental data, it also has an impressive
ability to explain the notable and unusual properties of the Bi
nanolines, such as the straightness (high kinking energy) and the
defect repulsion (DEZ). The large tensile strain in one direction and
zero strain in the other is very similar to the situation seen in the
silicide nanowires\cite{Chen2000}, and accounts for the invariant
width and extreme length of the nanowire.  Also of note are the
5-membered and 7-membered rings of Si, marked in Fig.~\ref{fig:struc}
(b) and (c). Such odd-membered rings are also present in the proposed
As-Ge step structure, suggesting that this structural motif may be of
general interest in structures involving Group V layers on Group IV
surfaces, particularly in situations where there is either compressive
or tensile stress.  Finally, the properties of these nanolines
demonstrate the importance of surface stress in the formation and
phenomenonology of nanoscale structures.

\begin{acknowledgments}
The authors would like to thank Prof. Andrew Briggs(Oxford University)
for useful discussions, and Bill McMahon for help with understanding
the As-Ge structures. We are also happy to acknowledge Dr. Masaru
Shimomura and Prof. Shozo Kono(Tohoku University) for detailed
discussions about their XPD experiments.  DRB acknowledges support
from the EPSRC and from the Royal Society through fellowships, and the
HiPerSPACE Centre (UCL) for computer time for DFT calculations. JHGO
is supported by the Japanese Science and Technology Agency (JST) as an
STA Fellow.  This study was performed through Special Coordination
Funds of the Ministry of Education, Culture, Sports, Science and
Technology of the Japanese Government (Research Project on active
atom-wire interconnects).
\end{acknowledgments}

\bibliography{BiLine}

\end{document}